\documentclass[conference]{IEEEtran}
\IEEEoverridecommandlockouts
\usepackage{cite}
\usepackage{amsmath,amssymb,amsfonts}
\DeclareMathOperator*{\argmax}{arg\,max}
\DeclareMathOperator*{\argmin}{arg\,min}
\usepackage{algorithm}
\usepackage{algpseudocode}
\usepackage{graphicx}
\usepackage{textcomp}
\usepackage{xcolor}
\usepackage{subfigure}

\def\BibTeX{{\rm B\kern-.05em{\sc i\kern-.025em b}\kern-.08em
    T\kern-.1667em\lower.7ex\hbox{E}\kern-.125emX}}
\begin{document}

\title{Beam Codebook Refinement for mmWave Devices with Random Orientations: Concept and Experimental Validation
%\thanks{This paper has been supported by the THULAB project established within T{\"U}B{\.I}TAK B{\.I}LGEM. The authors would like to thank all the THULAB researchers for their insightful and inspiring comments.}
}

\author{\IEEEauthorblockN{Bora Bozkurt\IEEEauthorrefmark{1}, Ahmet Muaz Aktaş\IEEEauthorrefmark{1}\IEEEauthorrefmark{2}, Hasan Atalay Günel\IEEEauthorrefmark{1}, Mohaned Chraiti\IEEEauthorrefmark{3}, Ali Görçin\IEEEauthorrefmark{1}\IEEEauthorrefmark{2}, İbrahim Hökelek\IEEEauthorrefmark{1}\IEEEauthorrefmark{2}}

\IEEEauthorblockN{\IEEEauthorrefmark{1} Communications and Signal Processing Research (HİSAR) Lab., T{\"{U}}B{\.{I}}TAK B{\.{I}}LGEM, Kocaeli, Turkey} 

\IEEEauthorblockN{\IEEEauthorrefmark{2}Department of Electronics and Telecommunications Engineering
Istanbul Technical University, Istanbul, Turkey}

\IEEEauthorblockN{\IEEEauthorrefmark{3} Department of Electronics Engineering, Sabanci University, {\.{I}}stanbul, Turkey} 

Emails: \it bora.bozkurt@tubitak.gov.tr, \it
    muaz.aktas@tubitak.gov.tr, \it
    hasan.gunel@tubitak.gov.tr, \\\it  
    mohaned.chraiti@sabanciuniv.edu, \it  ali.gorcin@tubitak.gov.tr and \it ibrahim.hokelek@tubitak.gov.tr
}

\maketitle

\begin{abstract}

There is a growing interest in codebook-based beam-steering for millimeter-wave (mmWave) systems due to its potential for low complexity and rapid beam search. A key focus of recent research has been the design of codebooks that strike a trade-off between achievable gain and codebook size, which directly impacts beam search time. Statistical approaches have shown promise by leveraging the likelihood that certain beam directions (equivalently, sets of phase-shifter configurations) are more probable than others. Such approaches are shown to be valid for static, non-rotating transmission stations such as base stations. However, for the case of user terminals that are constantly changing orientation, the possible phase-shifter configurations become equally probable, rendering statistical methods less relevant. On the other hand, user terminals come with a large number of possible steering vector configurations, which can span up to six orders of magnitude. Therefore, efficient solutions to reduce the codebook size (set of possible steering vectors) without compromising array gain are needed. We address this challenge by proposing a novel and practical codebook refinement technique, aiming to reduce the codebook size while maintaining array gain within $\gamma$ dB of the maximum achievable gain at any random orientation of the user terminal. We project that a steering vector at a given angle could effectively cover adjacent angles with a small gain loss compared to the maximum achievable gain. We demonstrate experimentally that it is possible to reduce the codebook size from $1024^{16}$ to just a few configurations (e.g., less than ten), covering all angles while maintaining the gain within $\gamma=3$ dB of the maximum achievable gain. 

\end{abstract}


\section{Introduction}
% SECOND 
The  demand for gigabit/s wireless links and massive connectivity is on the rise. Accordingly, the wireless industry is keen to explore millimeter-wave (mmWave) and higher frequencies, offering large bandwidth in the order of the gigahertz. Transmission over mmWave bands, however, comes at the expense of the sensitivity of transmissions to blockage and severe path loss. To compensate for the path loss, deploying large antenna arrays at the users' terminals is deemed necessary, leveraging the antenna array gain \cite{milimeter2018hemadeh} through beamforming which in general increase system level complexity. Consequently, the efforts are concentrated to develop novel beamforming approaches, in particular, to reduce the beam search time and increase achievable gain \cite{hierarchical2016xiao, multi2017noh, enhanced2018xiao}.

\subsection{Related Work}\label{secIA}
% SECOND

The high cost and power consumption of mmWave radio-frequency (RF) chains favor transceiver designs with a small number of RF chains and large antenna array \cite{milimeter2018hemadeh, beamforming2016kutty}. Consequently, analog beamforming (referred to here as beam-steering) emerged as a viable approach, in which the beam is steered by adjusting phase shifters, utilizing as few as one RF chain \cite{spatially2014ayach, channel2014alkhateeb, statistically2020shaban}.

Numerous efforts have been made to develop innovative approaches to devise the phase shifter configurations that maximize array gain while minimizing beam search time and overhead. These methods can be classified into two categories: channel state information (CSI)-based beam-steering \cite{spatially2014ayach,channel2014alkhateeb} and codebook-based beam-steering \cite{hierarchical2016xiao, multi2017noh, enhanced2018xiao, statistical2017khormuji, statisticsaided2018lin, datadriven2022ozkoc}. The CSI-based approach resembles its counterpart in digital beamforming, as it necessitates estimating the CSI for each antenna. The key distinction, however, lies in selecting the antenna configuration from a discrete set of potential phase-shifter configurations. Under the assumption of perfect CSI knowledge, the authors in \cite{spatially2014ayach,channel2014alkhateeb} showed that near-maximum array gain can be achieved. 

In \cite{spatially2014ayach}, the authors frame the beamforming precoder as a sparse reconstruction problem. They introduce a low-complexity algorithm for large antenna arrays, drawing inspiration from the well-known orthogonal matching pursuit technique. The work in \cite{channel2014alkhateeb} relies on the sparsity of the mmWave channel matrix to develop a low complexity channel estimation algorithm for beamforming purposes. Although these techniques demonstrate optimal performance, they rely on the assumption of having CSI knowledge for each antenna array element. However, factors such as the short channel coherence time (resulting in rapid channel variation), high path loss, and the deployment of a large number of antennas necessitate high pilot symbol rates to obtain viable CSI estimates \cite{milimeter2018hemadeh, beamforming2016kutty}. For instance, transmitting over the 30\,GHz bands may reduce the channel coherence time by a factor of 15 compared to transmitting over the 2\,GHz bands, suggesting an increase in the overhead of at least 15 times.

To obviate the need for CSI knowledge, various codebook-based beam-steering techniques have been proposed \cite{codebook2022mabrouki,hierarchical2016xiao, multi2017noh, enhanced2018xiao,statistically2020shaban, fast2018filippini, mmwave2019wang, statisticsaided2018lin, statistical2017khormuji, datadriven2022ozkoc, beamcodebook2019mo}. The main idea consists of steering a beam according to a set of predefined directions or configurations, which we refer here to as "codebook", and then selecting the one that maximizes array gain. Exhaustive search over all possible configurations is the most straightforward approach, yet the most time-consuming. To reduce the beam search time, also known as beam training, hierarchical codebook schemes have been explored in \cite{hierarchical2016xiao, multi2017noh}. These schemes employ a multi-layered codebook to progressively refine beamwidth through successive stages. This hierarchical approach conducts a binary search for the optimal beam, beginning with the low-resolution (coarse) beams and proceeding to the high-resolution (narrow) beams. Even though the hierarchical codebook scheme can present a time-efficient solution, its effectiveness is hindered by the high probability of selecting the direction pointing to a non-dominant multi-path component, resulting in high gain gap from the maximum \cite{hierarchical2016xiao}. Moreover, there is no solution to maintain the array gain close to the maximum gain.

To further reduce the search time and enhance array gain to be closer to the maximal gain, several codebook-based techniques assisted by statistical data have been put forward \cite{fast2018filippini, mmwave2019wang, statisticsaided2018lin, statistical2017khormuji, datadriven2022ozkoc}. The authors of \cite{statisticsaided2018lin} suggest collecting statistics about the Angle-of-Departure (AoD), assuming that certain angles are more probable than others. They then utilize the Vector Quantization technique to devise a set of AoDs that represent the entire set. The work in \cite{statistical2017khormuji} considers the scenario of multiple-antenna BS and single-antenna users. Consistent with the approach in \cite{statisticsaided2018lin}, the authors assume that the AoDs are not uniformly distributed. They subsequently develop an AoD quantization technique and a non-uniform codebook, in which certain codewords are more probable than others. The authors propose allocating power according to the probability of AoD to maximize coverage. The work in \cite{datadriven2022ozkoc} investigates the potential construction of a data-driven codebook that maximizes coverage probability, considering a multiple-antenna BS, single-antenna receivers and a predetermined codebook size. The algorithm iteratively adds new beam-steering codewords (phase shifters' configurations) to incrementally accommodate a new set of users until the pre-determined codebook size is reached.

\subsection{Problem Statement}
% SECOND 
Statistics-based codebooks offer a promising avenue for relaxing the requirement of CSI for beamforming and have shown superior performance compared to hierarchical techniques in terms of array gain. However, the existing approaches primarily focus on the BS with the goal of enhancing its coverage and average transmission rate, considering uneven spatial distribution of user terminals. They rely on the assumption that certain phase-shifters' configurations are more probable than others at the BS, a hypothesis that is less probable to occur at users' terminals due to their random rotations and the obstacles that affect the phase shifters' configurations. Developing a codebook refinement for the user terminals with random orientations remains an open problem, as mentioned in one of the recent statistics-based codebook design works \cite{datadriven2022ozkoc}.

In theory, the phase shifter configurations (steering vectors) within the codebook for user terminals with random orientations are expected to be uniformly distributed. To elaborate more, consider the case of an indoor office environment with an access node positioned at one end of a hallway. The access node chooses from the phase-shifters' configurations that direct the beam toward the hallway, whether the user is located in an office or in the hallway. In contrast, the user terminal's phase shifters' configuration may change according to the user's orientation and the surrounding obstacles. 

New user equipment, on the other hand, come with a large number of potential configurations. For example, the experimental devices used in this study have $1024^{16}$ possible steering vectors, making exhaustive search impractical. This naturally leads to the question: can the set of all possible configurations be reduced (codebook refinement) without compromising much on array gain?  

\subsection{Contributions}

In this paper, we consider the problem of codebook refinement at the user terminal with random orientations. The user terminal orientation may change randomly over time. There is no doubt that achieving optimal array gain at user terminals with random orientations requires scanning the entire codebook or CSI estimation. However, our objective is to investigate whether the codebook size can be significantly reduced by relaxing the requirement for optimal gain to maintaining array gain within a $\gamma$ decibel (dB) gap from the maximum gain. We project that a steering vector at a given angle could effectively cover adjacent angles (close steering vectors) with minimal gain loss compared to the maximum achievable gain. We collect a large offline dataset, and we use the proposed algorithm to identify the set of steering vectors that can cover all angles while maintaining the gap to the maximum within $\gamma$\,dB.

To validate the practicality aspect of the proposed approach, we conduct a real-world experiment using a 16-antenna receiver with an initial codebook size of $1024^{16}$, i.e., each antenna is armed with a 10-bit phase shifter. The refinement process takes $\gamma$, the target maximum allowed gap to the maximal gain, as input, and provides a smaller size codebook. Experimental results demonstrate that the proposed algorithm can reduce the codebook to fewer than 10 elements while maintaining a gap to the maximum gain as small as 3\,dB. This suggests that the approach strikes a good trade-off between beam search time (which is proportional to codebook size) and achievable array gain. We compare the performance of the proposed approach to the maximum gain as well as the one of hierarchical beam steering \cite{hierarchical2016xiao}.

The rest of this paper is organized as follows. Section II describes the system model. The proposed codebook refinement algorithm along with details regarding the codebook validation step is presented in Section III. The experimental setup, which includes a $4 \times 4$ Uniform Rectangular Array (URA), is detailed in Section IV. The measurement results are discussed in Section V. Finally, the paper is concluded in Section VI.

\section{System Model}
\label{sec:sysmodel} 
We consider a user terminal with an $L \times L$-antenna URA connected to a single RF chain. The terminal's azimuthal orientation, denoted by $\theta$, is assumed to be random and drawn from a uniform distribution. We assume that $\theta$ is unknown to the user terminal.
Each antenna is equipped with an $N$-bit configuration, {\it i.e.,} there are $2^N$ potential configurations per antenna. Among the $N$ bits, $K$ bits are allocated to control the amplitude of each antenna. Consequently, there are $2^{(N-K)}$ possible phase configurations per antenna. The total size of the codebook (set of possible steering vectors) is thus $2^{N\times L\times L}$. Without loss of generality, we assume an isotropic transmitter antenna. The received signal can be written as
\begin{equation}
        y = \boldsymbol{\hat{b}\,h}x + n.
\end{equation}
Here, $x \in \mathbb{C}$ represents the transmitted signal, while $\boldsymbol{h} \in \mathbb{C}^{1 \times N_{r}}$ denotes the complex channel vector between the base station and the user terminal. $\boldsymbol{\hat{b}} \in \mathbb{C}^{N_{r} \times 1}$ refers to the beam-steering phase shifters' configuration at the user terminal. Lastly, $n$ denotes the additive noise. The channel coefficients vector $\boldsymbol{h}$ can be expressed as follows,
\begin{equation}
    \boldsymbol{h} = \left[ \alpha_1 e^{-j\phi_{1}}, \alpha_2e^{-j\phi_2}, \ldots, \alpha_{L^2}e^{-j\phi_{L^2}} \right]^T
\end{equation}
where $\alpha_i$ and $\phi_i$ are the amplitude and phase of the channel coefficient at the $i^\text{th}$ antenna, respectively. 

Each antenna has the capability to undergo phase and amplitude configurations within a predefined grid (codebook). The resulting steering vector can be expressed as:
\begin{equation}
    \boldsymbol{b} = \left[ \hat\alpha_1 e^{-j\hat\phi_{1}}, \hat\alpha_2e^{-j\hat\phi_2}, \ldots, \hat\alpha_{L^2}e^{-j\hat\phi_{L^2}} \right]^T
\end{equation}
where \[\hat\alpha_i \in \left\{\frac{1}{2^K},\frac{2}{2^K},\dots,1\right\}\] and \[\hat\phi_i \in \left\{\frac{\pi}{2^{N-K-1}},\frac{2\pi}{2^{N-K-1}},\dots,2\pi\right\}\] for all $i\in 1..L^2$. Recall that $K$ refers to the number of bits that control the gain per antenna.

\section{Codebook Refinement}
\label{sec:coderef}

Let us denote the initial/unrefined codebook as $\mathcal{C}$, which encompasses all possible complex steering vectors (phase-shifter configurations), i.e., $\mathcal{C} = \{\boldsymbol{b}_i : i \in [1, |\mathcal{C}|]\}$ where $|\mathcal{C}| = 2^{N \times L \times L}$ is the size of the initial codebook. The codebook refinement process aims to identify a subset $\mathcal{\zeta} \subseteq \mathcal{C}$, with the smallest possible size $|\mathcal{\zeta}|$, that keeps the array gain within $\gamma$\,dB of the maximum array gain for any orientation $\theta \in \Theta$ of the user terminal.

The maximum received power obtained through scanning the refined codebook $\mathcal{\zeta}$, at a given user terminal orientation $\theta$, is denoted by $P_{\mathcal{\zeta}, \theta}$. For a given angle $\theta$, we use the maximum achievable received power $P_{\max, \theta}$ as benchmark. The achievable to the maximum array gain gap is expressed $$P_{\max,\theta}-P_{\mathcal \zeta,\theta}.$$
More formally, the codebook refinement process can be formulated as the following optimization problem:   
\begin{equation}
\label{cbproblem}
\begin{aligned}
\argmin_{\mathcal\zeta\, \subseteq \,\mathcal C} \quad & |\mathcal{\zeta}|\\
\textrm{subject to} \quad & P_{\max,\theta}-P_{\mathcal \zeta,\theta} \leq \gamma \text{ for all } \theta \in \Theta
\end{aligned}
\end{equation}
where $\Theta \subset [0,2\pi)$ is the set of all possible azimuthal receiver orientations.
The total number of possible subsets of the set \( \mathcal{C} \) is given by
\[ \sum_{i=1}^{|\mathcal{C}|} \binom{|\mathcal{C}|}{i} = 2^{|\mathcal{C}|} = 2^{2^{N \times L \times L}}. \] 

Although the refinement of the codebook is performed offline, the cardinality of the search space \( 2^{2^{N \times L \times L}} \) is nearly infinite, even for moderate values of \( N \) and \( L \) as considered in the experimental setup. Therefore, deriving \( P_{\text{max}, \theta} \) and its associated steering vector may necessitate consulting the entire codebook $\mathcal C$, a task that cannot be completed within reasonable time-frame due to the extensive size of the codebook prior refinement. As an alternative, and given the offline nature of codebook refinement, we rely on channel estimate-based beam-steering \cite{spatially2014ayach,channel2014alkhateeb}.\footnote{While CSI-based beam steering can be suitable during the offline codebook refinement phase, it may not be efficient for beam training, as discussed in Sec.\ref{secIA}.} This method achieves the optimal gain in case where the phase shifter configurations are continuous or finely granular \cite{spatially2014ayach}. For example, in the considered experiment, we have \( N = 10 \) bits suggesting 1024 possible configurations per antenna. We round the continuous steering vector, obtained through the Maximum Ratio Combiner (MRC), to the closest discrete vector in the codebook $\mathcal C$. For each antenna, the process of finding the phase-shifter configuration entails matching the corresponding element of the MRC steering vector, representing the ideal configuration for an antenna, to the nearest possible configuration.

The core of the proposed algorithm involves using measurements to iteratively construct the codebook $\mathcal{\zeta}$ with a selected set of steering vectors. We consider a large set of \(M\) random orientations $\Theta_{R} = \{\theta_m\}_{m=1}^{M}$. Initially, the codebook is empty. We select a random angle \(\hat{\theta}\) from $\Theta_{R}$, find its optimal steering vector \(\boldsymbol{\hat{b}}\), and add \(\boldsymbol{\hat{b}}\) to $\mathcal{\zeta}$. Next, for each remaining orientations $\theta$ in $\bar{\Theta}_R = \Theta_{R} \setminus \{\hat{\theta}\}$, we compare $P_{\mathcal \zeta,\theta}$ (the maximum received powers considering the codebook $\mathcal{\zeta}$) to $P_{\max,\theta}$. Let us denote by $\Theta_{\mathcal{\zeta}}$ the set of orientations for which the current version of the codebook satisfies the constraint on achievable power, i.e., $\theta\in \bar{\Theta}_R$ as such $P_{\max,\theta}-P_{\mathcal \zeta,\theta}\leq \gamma$\,dB. The set of remaining unsatisfied orientations is then updated as $\bar{\Theta}_R = \Theta_{R} \setminus \Theta_{\mathcal{\zeta}}$. The following operations are performed in each iteration: randomly select an angle from $\bar{\Theta}_R$, find the optimal steering vector, add it to the codebook $\mathcal{\zeta}$, update $\Theta_{\mathcal{\zeta}}$, and then update $\bar{\Theta}_R$. These operations are repeated for multiple iterations until $\bar{\Theta}_R$ becomes empty, i.e., $\Theta_{\mathcal{\zeta}}=\Theta_R$. For the sake of clarity, we outline the pseudocode for the refinement in the following algorithm.
\begin{algorithm}[H]
\caption{Codebook Refinement}\label{alg:refalgo}
\begin{algorithmic}
\State $\mathcal{\zeta} \gets \varnothing$
\State $\gamma$ \Comment{Set the array gain constraint threshold}
\State $\Theta_{\mathrm R},\,\, \Theta_{\mathcal \zeta}  \leftarrow \text{random.sample}(\Theta, M)$ %\Comment{Choose the set of orientations to apply refinement on}
\State $\Bar{\Theta}_{\mathrm R} \gets \Theta_{\mathrm R}$ %\Comment{$\Bar{\Theta}$ is initialized as $\Theta$}
%\State $i \gets 1$
\While{$\Bar{\Theta}_{\mathrm R} \neq \varnothing$}
\State $\hat{\theta} \leftarrow \text{random.sample}(\bar{\Theta}_{\mathrm R}, 1)$
\State $\boldsymbol{\hat{b}}=\underset{\boldsymbol{b}\in \mathcal{C}}{\argmax} \,\,\,P_{\hat\theta,\boldsymbol{b}}$
%\Bar{\Theta}$ \Comment{Pick a random angle from $\Bar{\Theta}$}
\State $\mathcal{\zeta} \gets \mathcal{\zeta} \cup \{\boldsymbol{\hat{b}}\}$ %\Comment{Generate the optimal steering vector for $\theta_i$ and add it to $\mathcal{\zeta}$}
\For{all $\theta \in \Bar{\Theta}_{\mathrm R}$}
\If{$P_{\text{max},\theta} - P_{\mathcal \zeta,\theta} \leq \gamma$}
\State $\Theta_{\mathcal \zeta} \gets \Theta_{\mathcal \zeta}\cup \{\theta\}$ %\Comment{Remove the angles that satisfy the condition}
\EndIf
\EndFor 
\State $\bar{\Theta}_{\mathrm R} \gets \bar\Theta_{\mathrm R} \setminus \Theta_{\mathcal \zeta}$
\EndWhile
\end{algorithmic}
\end{algorithm}

For codebook validation, we consider \( K \) random orientations. \( P_{\zeta} \) represents the maximum received power using all steering vectors in $\mathcal{\zeta}$. We compare \( P_{\zeta} \) to \( P_{\max} \), which is obtained through channel estimation based beam-steering. In the experiment, we consider \( M = 300 \) measurement points and \( K = 140 \) validation points for different $\gamma$ values. During the validation process, we randomly introduce obstacles between the transmitter and receiver. These obstacles vary in shape, and location, affecting the path-loss and the angle of arrival. For instance, a laptop screen placed close to the transmitter or receiver can cause an approximate loss of 30\,dB.

\section{Experimental Setup}\label{Subsec:exp}
A single antenna transmitter is placed at a distance of 5 meters from the $4\times4$ URA receiver module of EVK02004 \cite{evk02004} by Sivers Semiconductors. The distance between the Sivers antennas used is 5.15 millimeters (mm), and each of the 16 antennas is a square patch antenna with side lengths of 3.05\,mm. The Sivers URA is mounted on a rotating platform that goes from 45$^o$ to -45$^o$. Moreover, each antenna is equipped with a 10-bit phase shifter, providing 1024 possible configurations per antenna and resulting in a total codebook size of $1024^{16}$. The measurements are carried out in an indoor environment where there are many static but reflective objects. Moreover, during the validation, we place obstacles of varying sizes, materials, and shapes between the transmitter and the receiver.

\begin{figure}[]\vspace*{2pt}
    \centering
    \subfigure[Illustrative graph of the experimental setup ]{\includegraphics[width=3.5in]{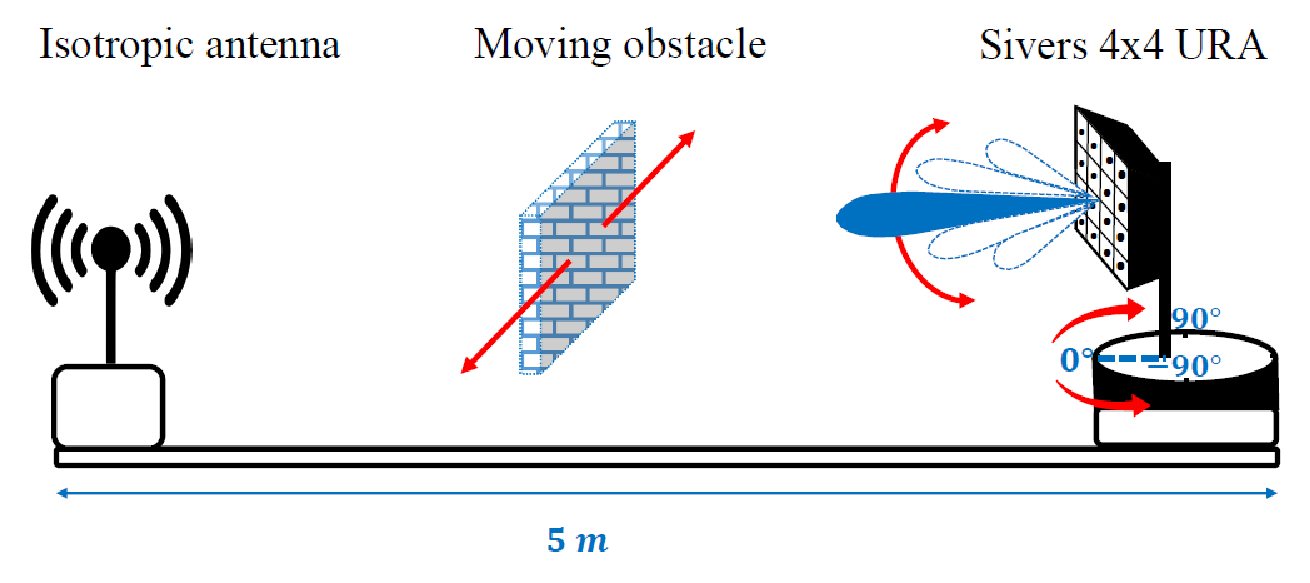}}
    \label{fig:illustration}
    
    \subfigure[Real setup]      {\includegraphics[width=3.5in]{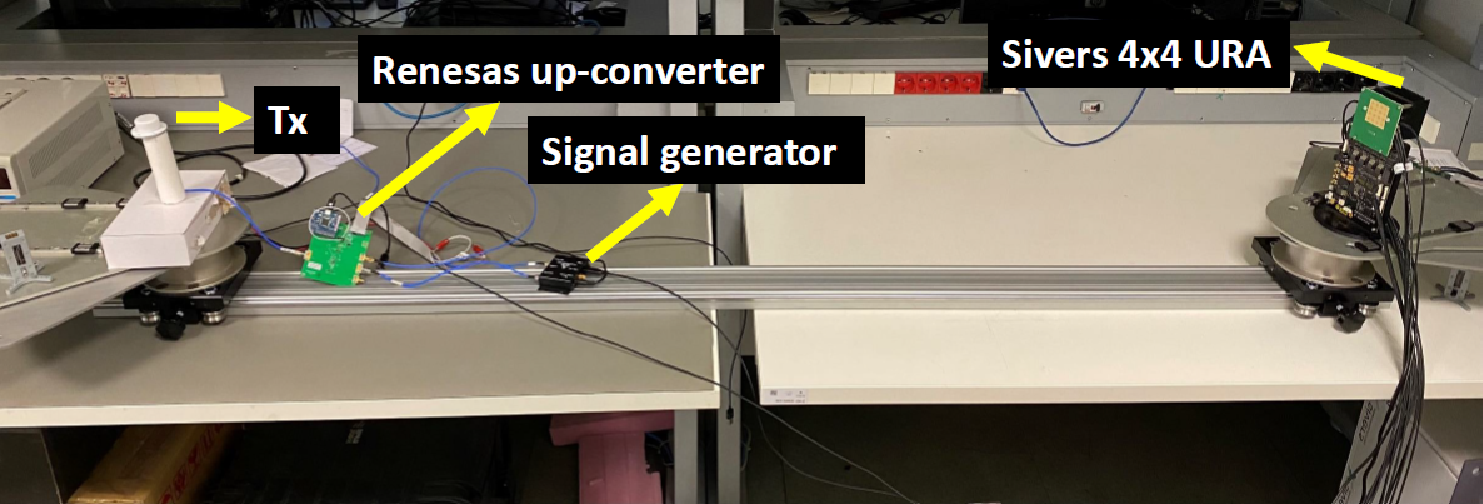}}
        \caption{\textbf{System model}}
    \label{exp setup}
\vspace{-0.3cm}
\end{figure}

To up-convert the carrier to 25.1 GHz, we use a Renesas F5728 frequency up-converter \cite{f5728} connected to a signal generator. The signal generator's primary function is to produce two continuous wave signals at the frequencies of 3.1 GHz and 5.5 GHz. The 5.5 GHz signal is connected to the local oscillator of the Renesas up-converter (shown in Fig. \ref{exp setup}. b), which includes a frequency multiplier by 4 to deliver a continuous wave signal with a frequency of 22 GHz. This output is then mixed with the 3.1 GHz signal, resulting in a signal with a carrier frequency of 25.1 GHz. This signal is fed to an isotropic antenna.

\par A computer is used to control the entire system, including the measurement collection and phase shifter configuration management. During the validation step following the codebook refinement, we rotate the device at a random azimuthal angle. We then steer a beam in various directions according to the phase-shifter configurations provided by the refined codebook $\mathcal{\zeta}$. Measurements are collected on the received power, and the optimal steering vector that maximizes the received power is identified through channel estimation. The gap to the maximum gain achieved is then measured and recorded.

\section{Experimental Results}
\label{subsec:results}

In this section, we report the results of a real-world measurement experiment to validate the proposed approach (see Sec. \ref{Subsec:exp}). For each random orientation $\theta$, we use the maximum received power $P_{\max,\theta}$ obtained through the channel estimate as benchmark. Moreover, we compare the performance of the proposed approach to that of the hierarchical beam-steering approach \cite{hierarchical2016xiao}. 

We consider $K = 140$ measurement points for validation. For each validation measurement point, we applied the following procedure: the multi-antenna receiver is oriented in a random direction $\theta$ and the refined codebook $\mathcal{\zeta}$ is scanned to identify the steering vector that maximizes the received power $P_{\mathcal \zeta,\theta}$. The performance of the proposed approach is assessed by analyzing the gap between the achievable and maximum array gain $P_{\max,\theta}-P_{\mathcal \zeta, \theta}$.
\begin{table}[h]
\centering
\hspace{-0.5cm}
\caption{Results for various $\gamma$ values}
\label{table1}
\begin{tabular}{|c|c|c|c|c|}
\hline
\textbf{Threshold $\gamma$} & 
\textbf {$| \mathcal{\zeta} |$} & \textbf{Number of samples} &
\textbf{Mean} & 
\textbf{Variance}\\
\hline
2 dB & 9 & 20 & 1.060 & 0.561\\
3 dB & 7 & 100 & 1.576 & 0.888\\
5 dB & 3 & 20 & 2.306 & 3.419\\
\hline
\end{tabular}
\end{table}

We analyze the performance of the proposed approach for different values of $\gamma$. Recall that the threshold $\gamma$ is set prior to the refinement stage, leading to different codebook sizes. In Table \ref{table1}, we present the mean and variance of the difference, \(P_{\max} - P_{\mathcal{\zeta}}\) along with the corresponding values of $\gamma$. The validation process can be time-consuming, requiring to perform channel based estimation beam-steering to find \(P_{\max}\) for each measurement point. Accordingly, we focus primarily on the case of $\gamma = 3$ dB, where we provide the results of over 100 validation measurements. However, we also analyze the other two cases with 20 measurements each. Table \ref{table1} demonstrates that, on average, the constraints on achievable power are satisfied. %In the following section, we provide further details on the performance of the proposed technique considerin

\begin{figure}[t]
\includegraphics[width = 1\linewidth]{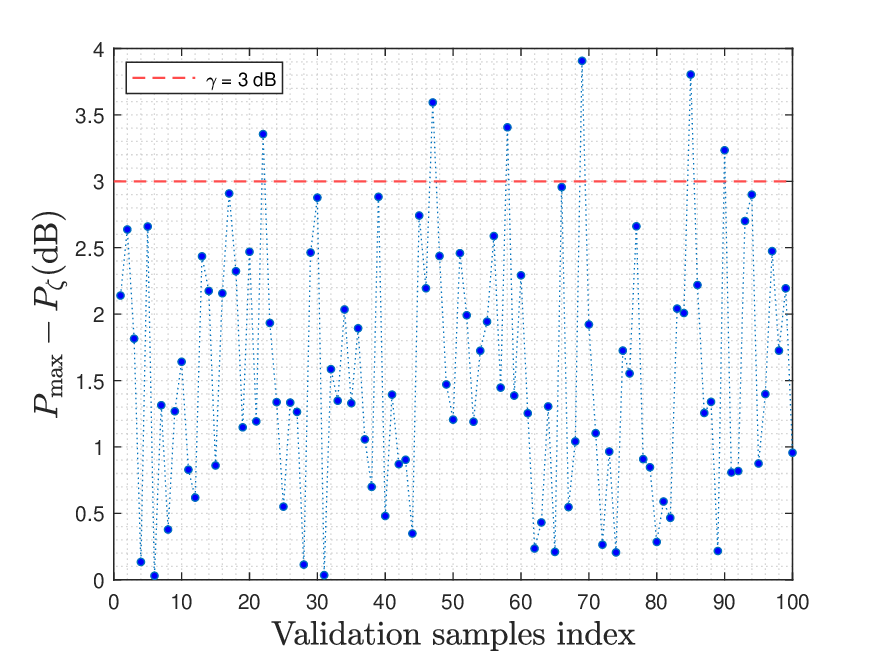}
\caption{The achievable to the maximum array gain for different samples }
\centering
\label{fig_power plot}
\end{figure}

Fig. \ref{fig_power plot} illustrates the gap between the maximum array gain and the achievable gain, \( P_{\max} - P_{\zeta} \), at each validation measurement point where \( \gamma \) is set to 3\,dB. The figure demonstrates that the vast majority of test measurements fall within the intended gain range. Notably, for only 6 data points out of 100, the achievable gain falls below the pre-set threshold. Despite this, all these points remain within 4\,dB of the maximum achievable power. While the refined codebook clearly provides performance close to that of the initial one, its size of 7 (as shown in Tab. II) represents a negligible fraction of the initial codebook size of \(2^{N \times L^2} = 1024^{16}\).

\begin{figure}[t]
\includegraphics[width = 1\linewidth]{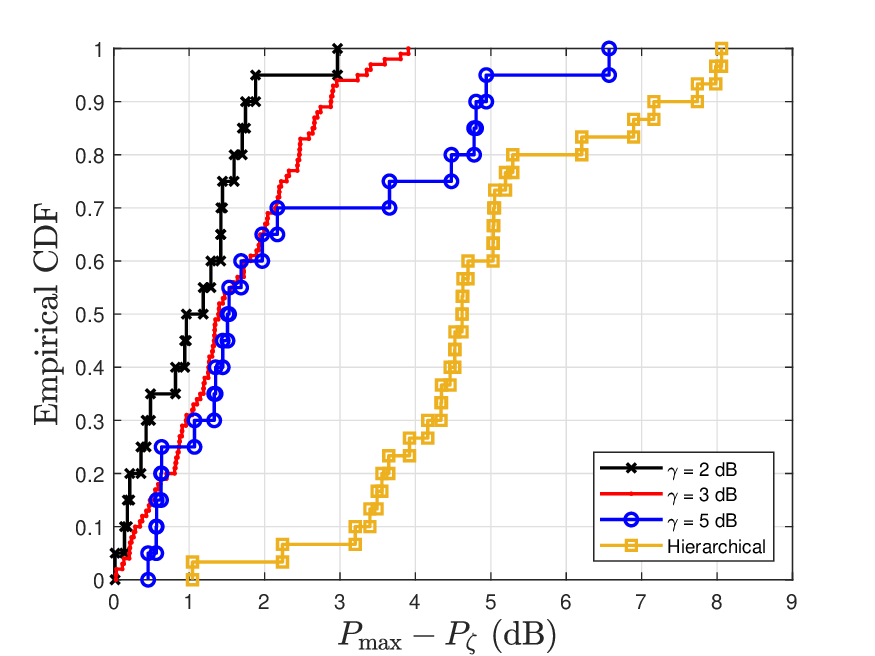}
\caption{Empirical CDF of the achievable to the maximum array gain}
\label{fig:emp cdf}
\centering
\end{figure}

To provide further insight into the efficiency of the proposed codebook refinement technique and the probability to maintain the gap within the intended gap, we provide the cumulative distribution function (CDF) of the achievable to maximum gain in Fig \ref{fig:emp cdf}. We present results for different intended values of $\gamma$. Recall that the threshold is set during the refinement stage, resulting in different codebook sizes. The results indicate that the constraint on achievable gain is satisfied for more than 90\% of cases. Moreover, with an additional 1 dB margin, the probability surpasses 97\% across all cases. In comparison to the gain achieved by a five-level hierarchical beam-steering technique, the proposed approach attains a gain closer to the maximum, even when considering a smaller codebook size. For example, with $\gamma$ set to 5 dB, the refined codebook consists of three entries, which is smaller than the five entries used in the five-level hierarchical beam-steering method. Nevertheless, the proposed approach achieves better average gain.

\section{Conclusion}
In this paper, we propose a data-driven codebook refinement framework for devices with random orientation. The experimental results demonstrate the possibility of reducing the codebook size from \(1024^{16}\) to fewer than a set of 10 potential steering vectors while maintaining the gain within a pre-set margin of \(\gamma\)\,dB compared to the maximum attainable gain (obtained via channel estimation). The results are validated through real-world measurements. The proposed approach outperforms its counterpart in the literature, namely the hierarchical beam-steering technique.

Despite the proven efficacy of the proposed technique in beam-steering, further investigation is required to develop a new approach for rapid beam alignment, particularly considering multiple-antenna transmitters and multiple-antenna receivers. Even when the codebooks at the transmitter and receiver are small (e.g., sizes \(N\) and \(K\) respectively), the total search space for beam alignment requires checking \(K \times N\) possibilities, which can still be significant. Reducing the search time for beam alignment will be the main focus of our future work. We aim to incorporate machine learning techniques to create a low-cost beam alignment approach.

\section{Acknowledgment}
This study has been carried out through the research vision of the THULAB project run at the  Informatics and Information Security Research Center (B{\.I}LGEM) of The Scientific and Technological Research Council of Türkiye (T{\"U}B{\.I}TAK).

This work has also received funding from the AIMS5.0 project. AIMS5.0 has been accepted for funding within the Key Digital Technologies Joint Undertaking (KDT JU), a public-private partnership in collaboration with the HORIZON Framework Programme and the national Authorities under grant agreement number 101112089.

\vspace{12pt}

\bibliographystyle{IEEEtran}
\bibliography{IEEEabrv,TWC1bib}

\end{document}